\documentclass[12pt]{iopart}

\usepackage{latexsym}
\usepackage{amssymb}

\usepackage{setstack}

\usepackage{bm, graphics}
\usepackage{graphicx}

\newcommand{\D}{\mbox{\rm d}}

\newcommand{\uh}[1]{\underline{\hat{#1}}}

\def\ket{\rangle}
\begin{document}

\title[Quantum-state input-output relations]{Quantum-state
input-output relations for absorbing cavities}

\author{M Khanbekyan\dag, D-G Welsch\dag, A A Semenov\ddag\P and W Vogel\ddag}

\address{\dag Theoretisch-Physikalisches Institut,
Friedrich-Schiller-Universit\"{a}t Jena, Max-Wien-Platz 1, D-07743
Jena, Germany}
\address{\dag Fachbereich Physik, Universit\"{a}t Rostock,
Universit\"{a}tsplatz 3, D-18051 Rostock, Germany}
\address{\P Institute of Physics, National Academy of Sciences of
Ukraine, 46 Prospect Nauky, UA-03028 Kiev, Ukraine}

\ead{mkh@tpi.uni-jena.de}

\date{\today}
\begin{abstract}
The quantized electromagnetic field inside and
outside an absorbing \mbox{high-$Q$} cavity
is studied,
with special emphasis on the absorption losses in the
coupling mirror and their influence on the outgoing field.
Generalized
operator input-output relations are
derived, which are used to calculate
the Wigner function of the
outgoing field.
To illustrate the theory, the preparation of the
outgoing field in a Schr\"{o}dinger cat-like
state is discussed.
\\[1ex]
{\bfseries Keywords:}
cavity QED, nonclassical states of the electromagnetic field, 
quantum state engineering, input-output relations
\end{abstract}


\maketitle

\section{Introduction}
\label{sec:intro}

Cavity QED provides a rich tool for
experimental tests and applications
of quantum mechanics, with special emphasis on
quantum decoherence and communication \cite{berman}.
The use of high-$Q$ cavities makes it possible
to control the evolution of coupled atom-field systems
in order to synthesize, at least on principle,
arbitrary nonclassical states of light
\cite{brattke:3534, kuhn:067901, raimond:565}.
For example, schemes have been developed
to generate various nonclassical states of light
by mapping atomic Zeeman states on
states of cavity fields
\cite{lange:063817, parkins:1578}.
Nonclassical states of light on their part
play a fundamental role in
quantum information processing
\cite{monroe:238, knill:46, pellizzari:3788}.

Unfortunately Nature imposes limits on us, because
of quantum decoherence, which is unavoidably connected with
the losses that are always present in practice.
In cavities both wanted and unwanted losses play a role. Whereas the
wanted losses, which result from the input-output coupling, typically
affect the quantum state of the cavity field, unwanted losses such as
scattering and absorption losses affect both the quantum state of
the cavity field and the quantum state of the outgoing field.
Keeping track of leaking emission
renders possible for
real-time measurement
of the quantum state of the field
and, therefore, offers the possibility of
feedback control of open cavities
\cite{doherty:012105},
provided that the input-output coupling can be regarded
as being the main source of decoherence.
For high-$Q$ cavities this must not necessarily be the case.
Moreover, when in quantum networks the radiation emitted by
cavities plays the role of
quantum communication channels
connecting, e.g., atoms trapped in distantly separated cavities
\cite{cirac:3221, browne:067901}, even small unwanted
losses can lead to a noticeable degradation of
the nonclassical features of the radiation
\cite{scheel:063811}.

In the rest of this article we
study the problem of the absorption losses of a high-$Q$
cavity and their effect on the quantum state of the outgoing field
in more detail, with special emphasis on the absorption losses
in the coupling mirror. Basing the calculations
on quantum electrodynamics in dispersing and absorbing media,
we derive a formula that relates the Wigner function of the outgoing
field to the Wigner functions of the cavity field,
the incoming field, and the dissipative channels.
As an application of the theory, we study
a scheme for generating Schr\"{o}dinger cat-like states
by combining two incoming modes prepared in coherent states
with a cavity mode prepared in a squeezed number state.

The paper is organized as follows. In section \ref{sec:2} the
cavity model is described and the basic equations are given.
Expressions for the cavity field and the outgoing
field are given in sections \ref{sec:3} and \ref{sec:4},
respectively. The Wigner function of the quantum state of the
outgoing field is derived in section \ref{sec:5}, and
section \ref{sec:6} presents an application.
Finally, the main results are summarized in section \ref{sec:7}.

\section{
Cavity model}
\label{sec:2}


\begin{figure}
\begin{center}
\includegraphics{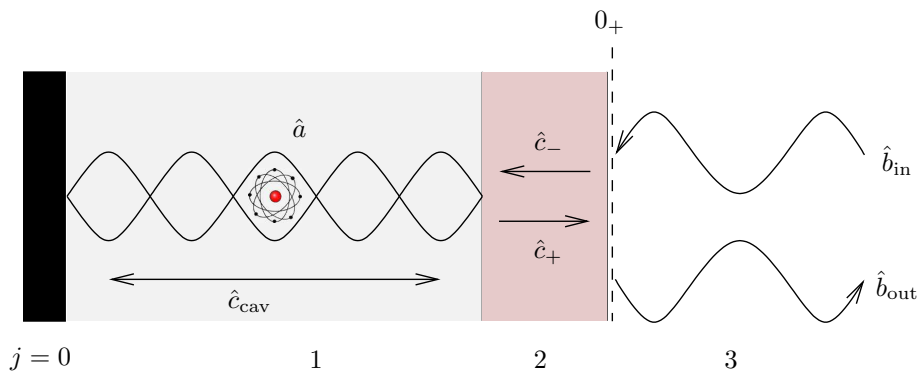}
\end{center}
\caption{\label{fig}Scheme of the cavity. The
fractionally transparent mirror (region (2)) is modelled by a
dielectric plate. The active sources are in region 1, which can
also contain some medium.}
\end{figure}
For simplicity let us consider a modified version of
Ley and Loudon's \cite{ley:227} cavity model,
defined by a planar multi-layer system according to
figure~\ref{fig},
where the layer labelled \mbox{$j$  $\!=$ $\!0$} is assumed
to be the totally reflecting mirror and the semitransparent mirror
(layer $2$) is modelled by a dielectric plate.
The linearly polarized electromagnetic waves propagate
in the $z$ direction, for which we use shifted coordinate systems
such that \mbox{$0$ $\!<$
$\!z$ $\!<$ $\!l$} for \mbox{$j$ $\!=$ $\!1$}, \mbox{$0$ $\!<$
$\!z$ $\!<$ $\!d$} for \mbox{$j$ $\!=$ $\!2$}, and \mbox{$0$ $\!<$
$\!z$ $\!<\infty$} for \mbox{$j$ $\!=$ $\!3$}.
Applying the one-dimensional version of the quantization scheme in
reference \cite{knoell:1, ho:053804},
allowing for active atomic sources inside the cavity
[$A$th atom with electric dipole moment $\hat{ d}_A (t)$ being at
position $z_A$],
and following reference \cite{khanbekyan:063812},
we may represent the electric field in the $j$th
layer of permittivity $\varepsilon_j(\omega)$ \mbox{($j$ $\!=$
$\!1,2,3$)} in the form of
\begin{equation}
\label{2.0}
\hat{E}^{(j)}(z,t) = \int_0^\infty \D\omega\,
\uh{ E}{^{(j)}}(z, \omega,t) + \mathrm{H.c.},
\end{equation}
\begin{equation}
\label{2.1}
\uh{ E}{^{(j)}}(z, \omega,t)
= \uh{ E}^{(j)}_\mathrm{free}(z, \omega,t)
+ \uh{ E}^{(j)}_\mathrm{s}(z, \omega,t),
\end{equation}
where
\begin{equation}
\label{2.1-1}
\uh{ E}^{(j)}_\mathrm{free}(z, \omega,t)
= i\omega\mu_0 \sum_{j'=1}^3\int_{[j']}\D z'\,
G^{(jj')}(z,z',\omega)\hat{\underline{j}}^{(j')}_\mathrm
{free}(z',\omega,t)
\end{equation}
is the free-field part
($[j']$ indicates integration over the $j'$th layer),
\begin{equation}
  \label{2.2}
\underline{\hat{ E}}^{(j)}_\mathrm{s} (z, \omega, t) =
    \frac{i}{\pi \epsilon _0
      \mathcal{A}}\,
    \frac{\omega ^2} {c^2}
    \!\sum_A\! \int\! \D t' \,\Theta(t-t')
   e^{-i\omega (t-t')}
   \hat{ d}_A (t')\, \mathrm{Im}\,G ^{(1j)}(z_A, z,\omega)
\end{equation}
($\mathcal{A}$: mirror area) is the source-field part, and
\begin{equation}
      \label{2.3}
      \uh { j}^{(j)}_\mathrm{free} (z, \omega,t ) =
      \omega\, \sqrt { \frac {\hbar \epsilon _0} {\pi {\cal A} }\,
      \varepsilon '' _j (\omega) } \,
      \hat { f}^{(j)}_\mathrm{free} (z, \omega,t )
\end{equation}
[$\hat { f}^{(j)}_\mathrm{free} (z, \omega,t )$ $\!=$
$\!e^{-i\omega(t-t')} \hat { f}^{(j)}_\mathrm{free}(z, \omega,t' )$;
$\hat { f}^{(j)}(z, \omega,t )$, bosonic fields; $\varepsilon_j(\omega)$
$\!=$ $\!\varepsilon'_j(\omega)$ $\!+$ $\!i\varepsilon_j''(\omega)$].
Further,
\begin{eqnarray}
      \label{2.5}
      G^{(jj')}(z, z', \omega )
      &=& {\textstyle\frac{1}{2}} i \bigl[
      \mathcal{E} ^{j>}    (z, \omega )\,
      \Xi^{jj'}    \mathcal{E} ^{j'<}    (z', \omega)
      \Theta (j-j')
\nonumber \\
&&
      + \mathcal{E}^{j<}    (z, \omega )\,
      \Xi^{j'j}   \mathcal{E} ^{j'>}   (z', \omega )
      \Theta (j'-j)
      \bigr]
\end{eqnarray}
is the (nonlocal part of the) Green function,
where the functions
\begin{equation}
      \label{2.7}
\mathcal{E} ^{(j)>}    (z, \omega ) =
       e^{i \beta _j (z-d _j)}
      + r  _{j/3}
      e^{-i \beta _j (z-d _j)}
      \end{equation}
and
\begin{equation}
      \label{2.9}
      {\mathcal{E}} ^{(j)<}    (z,   \omega )\, =\,
        e^{-i \beta _j z}
      + r  _{j/0}   e^{i \beta _j z}
\end{equation}
represent waves of unit strength travelling, respectively,
rightward and leftward in the $j$th layer and being reflected at
the boundary [note that $\Theta (j$ $\!-$ $\!j')$  means
\mbox{$\Theta (z$ $\!-$ $\!z')$} for $j$ $\!=$ $\!j'$]. Further,
$\Xi^{jj'}$ is defined by
\begin{equation}
      \label{2.11}
      \Xi^{jj'} =
      \frac{1}{\beta_3 t _{0/3}}\,
      \frac{t _{0/j}e^{ i \beta _j d _j}}{D _{ j}}\,
      \frac{t _{3/j'}e^{ i \beta _{j'} d _{j'}}}{D _{ j'}}\,,
      \end{equation}
where
\begin{equation}
      \label{2.13}
      D_{j} = 1 - r _{j/0} r _{j/3} e^{2 i \beta _j d _j}
      \end{equation}
and
\begin{equation}
      \label{2.15}
      \beta _j \equiv
      \beta _j(\omega) =
      \sqrt {\varepsilon _j (\omega)} \,\frac {\omega} {c}
      = \left[n_j ' (\omega) + i n_j '' (\omega) \right]\,\frac {\omega} {c}
      = \beta _j ' + i \beta _j ''
\end{equation}
[$\beta_j ', \beta_j ''$ $\! \geq$ $\! 0$,
$d_1$ $\!=$ $\!l$, $d_2$ $\!=$ $\!d$,
$d_3$ $\!=$ $\!0$].
The quantities \mbox{$t_{j/j'}$ $\!=$
$\!(\beta_j/\beta_{j'})t_{j'/j}$} and $r_{j/j'}$
are
respectively
the transmission and reflection coefficients between
the layers $j'$ and $j$, which can be recursively determined.
Note, for the cavity model under consideration
\mbox{$r_{10}$ $\!=$ $\!-1$}, since
perfect reflection from the left-side mirror
 has been assumed.

\section{Cavity field}
\label{sec:3}

To evaluate, according to equations (\ref{2.0})--(\ref{2.2})
for $j$ $\!=$ $\!1$, the electric field inside the cavity,
we notice that the function
$D_1(\omega)$ given by equation (\ref{2.13})
defines the spectral response of the cavity. That is,
the complex resonance frequencies
$\Omega_{k}$
$\!=$ $\!\omega_{k}$
$\!-$ $\!{\textstyle\frac {1} {2}}i\Gamma _{k}$
of the cavity field
are determined by the zeroes of
\begin{equation}
      \label{2.17}
 D_1 (\Omega _{k})
    = 1+ r_{13}(\Omega _{k})
    e^{2 i \beta_1 (\Omega _{k}) l} = 0
\end{equation}
and may be found by iteration. Somewhat lengthy
but straightforward calculations show that -- provided
that $\Gamma_k$ $\!\ll$
$\!\Delta\omega_k$, with \mbox{$\Delta\omega_k$ $\!=$
$\!\frac{1}{2} (\omega_{k+1}$ $\!-$ $\!\omega_{k-1})$} being the
width of the $k$th interval -- equation (\ref{2.0}) [together with
equations (\ref{2.1})--(\ref{2.2})] can be approximately rewritten
as
\begin{equation}
      \label{2.59}
      \hat{ E}^{(1)}(z, t) = \sum_k
      E_k(z)
      \hat{a}_k (t) + \mathrm{H.c.},
\end{equation}
where the standing wave mode functions are defined as
\begin{equation}
  \label{2.61}
  E_k (z) = i\omega _k
  \left[\frac {\hbar}
  { \epsilon_0 \varepsilon _1( \omega_k) l {\cal A} \omega_k}
  \right]^{\frac{1}{2}}\sin[\beta_1(\omega_k) z] ,
\end{equation}
and
\begin{eqnarray}
\label{2.63}
\hat{a}_k(t) =
        \int \D  t' \, \Theta ( t - t')
        e^{-i\Omega_{k}(t-t')}
\nonumber \\
\hspace{2ex}
\times
\biggl\{
\biggl[ \frac {c} {2 n_1(\omega_k) l}\biggr]^{\frac{1}{2}}
\biggl[
T_k \hat{b}_{k\mathrm{in}} ( t') +
  \sum _{\lambda }
    A _{k {\lambda }} \hat{c}_{k {\lambda }} ( t')
\biggr]
      -\frac{i}{ \hbar }
      \sum_A
      E_k (z_A)
      \hat{ d}_A (t')
  \biggr\}.
\end{eqnarray}
Here the operators $\hat{c}_{k \lambda}( t)$
and $\hat{b}_{k\mathrm{in}}( t)$ are defined by
\begin{eqnarray}
\label{2.51}
&        \hat{c}_{k \lambda}( t)
        = \frac{ 1}{\sqrt{2 \pi}}
        \int_{(\Delta_k)} \D\omega\,
        \hat{c}_{\lambda}(\omega, t),
\\
\label{2.53}
&        \hat{b}_{k\mathrm{in}}( t)
        = \frac{ 1}{\sqrt{2 \pi}}
        \int_{(\Delta_k)} \D\omega\,
        \hat{b}_{\mathrm{in}}(\omega, t),
\end{eqnarray}
where
the integration runs in the interval
\mbox{$(\Delta_k)$ $\!\equiv$
$[\frac{1}{2}(\omega_{k-1}$ $\!+$
$\!\omega_k),\frac{1}{2}(\omega_k$ $\!+$ $\!\omega_{k+1})]$},
and
the bosonic operators $\hat{b}_{\mathrm{in}}(\omega, t)$
and $\hat{c}_{\lambda}(\omega, t)$ ($\lambda$ $\!=$ $\!\mathrm{cav},\pm$)
read
\begin{eqnarray}
\label{2.31}
&
\hat{b}_{\mathrm{in}}(\omega, t) =
       -
       \sqrt{\frac{\mu_0c \pi\mathcal{A}}{\hbar\omega}}
    \frac{|n_3|}{n_3\sqrt{n_3'}}
     \int_{[3]} \D z'\,
    e^{i \beta _3 z'}
    \underline{\hat { j}}^{(3)}_\mathrm{free} (z', \omega, t)
,
\\
\label{2.33}
&    \hat{c}_{\pm} (\omega, t) =
    -
    \alpha _{\pm}
    \sqrt{\frac{\mu_0c \pi\mathcal{A}}{\hbar\omega}}
    \frac{ 1}{2 n_2}
     \int_{[2]} \D z'\,
    \left[
       e^{- i \beta _2 (z'-d)}
       \pm
        e^{ i \beta _2 z'}
        \right]
        \uh { j}^{(2)}_\mathrm{free} (z', \omega, t) ,
\\
\label{2.35}
&
\hat{c}_\mathrm{cav}(\omega, t) =
    -\alpha_\mathrm{cav} \sqrt{\frac{\mu _0 c \pi\mathcal{A}}{\hbar\omega}}
    \frac{ 1}{2 n_1}
    \int_{[1]} \D z\,
    \sin(\beta _1 z)
     \uh{ j}^{(1)}_\mathrm{free} (z,\omega, t)
,
\end{eqnarray}
with
\begin{eqnarray}
  \label{2.37}
&   \alpha_\mathrm{cav} = \alpha_\mathrm{cav} (\omega)
= 2\sqrt{2}|n_1|
    \left[n_1 '\sinh(2 \beta _1'' l)
    - n_1'' \sin(2 \beta _1 'l)\right]^{-\frac{1}{2}} ,
\\
\label{2.39}
&    \alpha_{\pm} = \alpha _{\pm}(\omega)
= |n_2|e^{ \beta _2'' d /2}
    \left[n_2'\sinh(\beta _2'' d)\pm n_2''\sin(\beta _2'd)
    \right]^{-\frac{1}{2}}
,
\end{eqnarray}
and the coefficients $T_k$, $A_{k\lambda}$ are defined
according to
\begin{eqnarray}
   \label{2.41}
T_k = T(\omega_k ),&\quad&
T(\omega )
=
- \frac{t_{31}\sqrt{n_1
n_3'} }{|n_3|}
  \,e^{
i \beta _1 l}
,
\\
  \label{2.43}
A_{k\pm} = A _{\pm}(\omega_k ),&\quad&
A _{\pm}(\omega )
=
  - \frac {t_{21}\sqrt{n_1
}} {D_2'\alpha_{\pm }}
  \left(r_{23} e^{i \beta _2 d} \pm 1\right)
  e^{
i \beta _1 l}
,
\\
  \label{2.45}
A _{k \mathrm{cav}} = A _{ \mathrm{cav}}(\omega_k ),&\quad&
A _{ \mathrm{cav}}(\omega )
=
-4 i \,\frac{\sqrt{n_1
}}{\alpha _\mathrm{cav}}
\,.
\end{eqnarray}

It can be proved that the operators $\hat{b}_{k\mathrm{in}}(t)$
and $\hat{c}_{k\lambda}(t)$ satisfy,
on a (course-grained) timescale \mbox{$\Delta t$ $\!\gg$
$\Delta\omega_k^{-1},\Delta\omega_{k'}^{-1}$},
the commutation relations
\begin{eqnarray}
\label{2.55}
&
\bigl[\hat{b}_{k\mathrm{in}}(t),
\hat{b}_{k'\mathrm{in}}^\dagger(t')\bigr]
= \delta_{kk'}\delta (t - t')
,
\\
\label{2.57}
&
\bigl[\hat{c}_{k\lambda}(t),
\hat{c}_{k'\lambda'}^\dagger(t')\bigr]
= \delta_{kk'}\delta_{\lambda\lambda'}\delta (t - t')
,
\end{eqnarray}
and
the equal-time commutation relation
\begin{equation}
  \label{2.64}
  [\hat{a}_k(t),\hat{a}_{k'}^{\dagger}(t)] = \delta_{kk'}
\end{equation}
hold. {F}rom equation~(\ref{2.63}) it follows that
$\hat{a}_k(t)$ obeys the Langevin equation
\begin{eqnarray}
    \label{2.65}
\dot{\hat{a}}_k
        &=& - i\left(\omega_{k}
        - {\textstyle\frac{1}{2}}i\Gamma_k\right) \hat{a}_k
        - \frac{i}{ \hbar}
        \sum_A
        E_k (z_A)
        \hat{ d}_A
\nonumber\\
&&
+ \biggl[ \frac {c} {2 n_1(\omega_k) l}\biggr]^{\frac{1}{2}}
\biggl[ T_k \hat{b}_{k\mathrm{in}} ( t)
+ \sum _{\lambda } A _{k\lambda } \hat{c}_{k \lambda } ( t)
\biggr]
.
\end{eqnarray}
Within the approximation scheme used,
the total
decay
rate $\Gamma_k$
determined from equation (\ref{2.17})
can be decomposed into a radiative part $\gamma_{k\mathrm{rad}}$ and a
non-radiative part $\gamma_{k\mathrm{abs}}$,
\begin{equation}
\label{2.67}
        \Gamma_k  =
         \gamma_{k\mathrm{rad}}
        +
        \gamma_{k\mathrm{abs}},
\end{equation}
where
\begin{eqnarray}
\label{2.68}
&     \gamma_{k\mathrm{rad}}
     = \frac{c}{2 |n_1(\omega_k)| l}   |T_k|^2,
\\
\label{2.69}
&     \gamma_{k\mathrm{abs}} =
    \sum _{\lambda }\gamma_{k\lambda} =
     \frac{c}{2 |n_1(\omega_k)| l}
    \sum _{\lambda }|A_{k\lambda }|^2
.
\end{eqnarray}
Note, that Langevin equations
of the type of equation (\ref{2.65}) can be also obtained from quantum
noise theories on the basis of an appropriately
chosen Senitzky-Gardiner-Collett Hamiltonian
\cite{senitzky:227,gardiner:1386}.

\section{Input-output relations}
\label{sec:4}

For simplicity we restrict ourselves to the case of a cavity in free
space, i.e., $n_3\to 1$.
To determine the input-output relations, we decompose the
field outside the cavity [equations(\ref{2.0})--(\ref{2.2})
for $j$ $\!=$ $\!3$
together with the Green function (\ref{2.5})]
into incoming and outgoing parts and introduce the bosonic operators
\begin{equation}
      \label{2.71}
      \hat{ b}_{ \mathrm{out}}  (\omega, t)
      =
      2
      \sqrt{\frac{\pi\mathcal{A}}{\mu_0c\hbar\omega}}
     \,\uh{ E}^{(3)} _{\mathrm{out}}(z, \omega, t)
     \bigr|_{z=0^+}
.
\end{equation}
Recalling the resonance structure of the spectral response of the
cavity, we may write
\begin{equation}
\label{2.75}
\hat{ b}_{\mathrm{out}}(t)
= \frac{1}{\sqrt{2\pi}} \int
\mathrm{d}\omega\,
\hat{ b}_{ \mathrm{out}}(\omega,t)
= \sum_k \hat{b}_{k\mathrm{out}}(t),
\end{equation}
\begin{equation}
\label{2.77}
\hat{ b}_{k\mathrm{out}}(t)
= \frac{1}{\sqrt{2\pi}} \int_{(\Delta_k)} \mathrm{d}\omega\,
\hat{ b}_{k\mathrm{out}}(\omega,t),
\end{equation}
where, within the approximation scheme used,
\begin{eqnarray}
  \label{2.79}
\fl
        \hat{b}_{k\mathrm {out}}(\omega, t)
        =
        \sqrt{\frac{\pi\mathcal{A}}{\mu_0c\hbar\omega}}
        \frac{
        \omega _k t_{13}\exp [i \beta _1(\Omega _k) l]}
     {\epsilon _0
       \varepsilon _1 \!(\Omega _k) l
       \mathcal{A}
     }
\nonumber\\ \hspace{-10ex} \times\,
    \sum_A \int\! \D t' \Theta(t\!-\!t')
   e^{-i\Omega_k (t-t')}
   \hat{ d}_A (t')
  \sin [\beta _1\!(\Omega _k) z_A]
\nonumber \\ \hspace{-6ex}
+ \frac{c}{2n_1 l}\,T
 ^{({\rm o})}(\omega)
        \int \D  t' \, \Theta ( t - t')
        e^{-i\Omega_{k}(t-t')}
     \biggl[
      T
      (\omega )
      \hat {b}_{
      \mathrm{in}}(\omega, t')
      + \sum _{\lambda }
      A_{
      \lambda}(\omega )
      \hat {c}_{
      \lambda}\!(\omega, t')
      \biggr]
\nonumber\\ \hspace{-6ex}
        + A _{
        +} ^{({\rm o})}(\omega)\hat{c}_{
        +}(\omega, t)
        + A _{
        -} ^{({\rm o})}(\omega)\hat{c}_{
        -}(\omega, t)
        + R
        ^{({\rm o})}(\omega)\hat{b}_{
         \mathrm{in}} (\omega, t )
,
\end{eqnarray}
with
\begin{eqnarray}
  \label{2.81}
&  A_{
  \pm} ^{({\rm o})}(\omega) =
  \frac {t_{23}} {D_2'}
  \frac{1\pm r_{21} e^{i \beta _2 d}}{\alpha_{\pm}}\,,
\\
  \label{2.83}
&  T
  ^{({\rm o})}(\omega) =
  \frac{ t_{13}} {\sqrt{n_1}}
\, e^{i \beta _1 l}
,
\\
  \label{2.85}
&  R
  ^{({\rm o})}(\omega ) =
r_{31}.
\end{eqnarray}

Performing in equation~(\ref{2.77}) the $\omega$ integration
and recalling
equations (\ref{2.63}), (\ref{2.51}), and (\ref{2.53}),
we obtain the generalized input-output relations
\begin{equation}
\label{2.87}
    \hat{b} _{k\mathrm {out}}(t)
    =
    \left[\frac{c}{2n_1(\omega_k) l}\right]^{\frac{1}{2}}
    \!T_k^{({\rm o})}\hat {a} _k(t)
    \!+\! R_k ^{({\rm o})}\hat{b} _{k\mathrm{in}}(t)
    \!+\! A _{k+} ^{({\rm o})}\hat{c}_{k+}(t)
    \!+\! A _{k-} ^{({\rm o})}\hat{c}_{k-}(t)
\end{equation}
[$T_k^{({\rm o})}$ $\!=$ $\!T^{({\rm o})}(\omega_k)$,
$A^{({\rm o})}_{k\pm}$ $\!=$ $\!A^{({\rm o})}_{\pm}(\omega_k)$,
$R_k^{({\rm o})}$ $\!=$ $\!R^{({\rm o})}(\omega_k)$].
In close analogy to equation (\ref{2.55}), the commutation relation
\begin{equation}
\label{2.89} \bigl[
        \hat {b} _{k{\rm out}}  ( t) ,
        \hat {b}^{\dagger} _{k'{\rm out}} ( t')
         \bigr]
         = \delta _{kk'} \delta (t-t ')
\end{equation}
can be shown to hold.
Omitting the last two terms on the right-hand side
in equation (\ref{2.87}), which
describe
the effect on the outgoing field of the
noise associated with the losses in the coupling mirror,
we recover the input-output relations suggested by standard
quantum noise theories \cite{gardiner:1386}.
Note, that a Hamiltonian of the Senitzky-Gardiner-Collett type
\cite{senitzky:227, gardiner:1386} is not suited for
consideration of the effect.

\section{Quantum-state extraction}
\label{sec:5}

Let us suppose that at some initial time $t_0$
the $k$th cavity mode is prepared
in a certain quantum state
and evolves freely for $t$ $\!\geq$ $t_0$
(the preparation time is assumed to be short
compared to the decay time $\Gamma _k ^{-1}$
of the cavity mode).
{F}rom equations (\ref{2.65}), (\ref{2.79}), and (\ref{2.87})
it then follows that
\begin{equation}
\label{3.23}
        \hat{b}_{k \mathrm {out}}
        (\omega, t) =
        F^*_k (\omega, t)
        \hat{a}_k(t_0)
        + \hat{B}_k(\omega,t),
\end{equation}
where
\begin{equation}
\label{3.25}
        F_k (\omega, t) =
         \frac{i} {\sqrt{2\pi}}
        \left(\frac{c}{2 n_1^*  l}\right)^{1/2} \!
        T_k^{({\rm o})*}
        f_k(\omega, t)
        ,
\end{equation}
\begin{equation}
  \label{3.27}
  \hat{B}_k(\omega,t)\! =
      \!\!\int _{(\Delta_k)}
       \!\D\omega '
\Bigl[
G_{k\rm{in}}^*
(\omega, \omega ', t)
       \hat{b}_{k\mathrm {in}} (\omega ', t_0)
\!+ \!\sum_{\lambda}
G_{k\lambda}^*
(\omega, \omega ', t)
 \hat{c} _{k\lambda} (\omega',t _0)
\Bigr]
,
\end{equation}
with
\begin{eqnarray}
  \label{3.34}
&   f_k(\omega, t)
   =
    \frac
         {\exp \left[-i (\omega -\Omega ^* _k) (t+\Delta t-t_0)
                \right] -
          1
         }
         {\omega - \Omega ^* _k}
   \,e^{i\omega  (t-t_0) }
,
\\
\label{3.35}
&       \upsilon (\omega, \omega ', t)  =
        \frac {1}{2\pi}\frac{c}{2n_1 ^* l}
        \frac{1}
        {\omega -\omega'  }
    [f(\omega, t) - f(\omega ' , t) e^{i(\omega '-\omega) \Delta t}]
,
\end{eqnarray}
\begin{eqnarray}
  \label{3.29}
&
G_{k\rm{in}}
        (\omega, \omega ', t) &=
        T_k ^{({\rm o})*} T_k^*
        \upsilon _k (\omega, \omega ', t)
        + R_k ^{({\rm o})*}
         e^{i \omega ' (t-t_0)}
        \delta (\omega - \omega '),
\\
\label{3.31}
&
G_{k\rm{cav}}
         (\omega, \omega ', t) &=
        T_k ^{({\rm o})*} A_{k \mathrm{cav}}^*
        \upsilon_k (\omega, \omega ', t),
\\
\label{3.33}
&
G_{k\pm}        (\omega, \omega ', t) &=
        T_k ^{({\rm o})*} A^*_{k\pm}
        \upsilon _k (\omega, \omega ', t)
        + A^{({\rm o})*}_{k\pm}
         e^{i \omega ' (t-t_0)}
        \delta (\omega - \omega ')
.
\end{eqnarray}
For the following it will be convenient to
introduce unitary, explicitly time-dependent
transformations according to
\begin{eqnarray}
  \label{6.3}
\hat{b}_{
    k
      \mathrm {out}} (\omega , t)
&=
      \sum_i
      \phi _{\mathrm {out}}^{(i)*}
    (\omega
     ,t
      )
    \hat{b}_{k\mathrm {out}}
        ^{(i)} (t)
,
\\
  \label{6.4}
    \hat{b}_{
    k
      \mathrm {in}} (\omega , t_0)
&=
      \sum_i
      \phi _{\mathrm {in}}^{(i)*}
    (\omega
     ,t_0
      )
    \hat{b}_{k\mathrm {in}}
        ^{(i)} (t_0)
,
\\
  \label{6.5}
      \hat{c}_{
        k\lambda} (\omega , t_0)
      &=
      \sum_i
      \phi _{\lambda} ^{(i)*}
     (\omega
     ,t_0
      )
       \hat{c}_{k\lambda}
       ^{(i)} (t_0)
,
\end{eqnarray}
where, for chosen $t$ and $\varsigma$
($\varsigma$ $\! =$ $\!\mathrm{out},\mathrm{in},  \lambda$),
the functions $\phi_\varsigma^{(i)}(\omega,t)$
are complete sets of square integrable orthonormal functions.
Needless to say, that the bosonic commutation
relations are preserved.

{F}rom equation (\ref{3.23}) it is easy to see, that the cavity
mode predominantly escapes into the outgoing mode with
\begin{equation}
  \label{6.19}
   \phi_{\mathrm {out}} ^{(1)}  (\omega
   ,t
   ) = \frac
          {F_k(\omega , t)}
          {\eta _k(t)}\,,
\end{equation}
where
\begin{equation}
 \label{6.21}
         \eta_k(t)  =
         \left[
         \int _
         {(\Delta_k)}
         \D\omega\, |F_k(\omega ,t)| ^2
         \right]^{1/2}
.
 \end{equation}
Hence,
for chosen $k$,
equation~(\ref{3.23}) takes the form
\begin{equation}
\label{6.23}
\hat{b}_{k\mathrm {out}}^{(i)} (t) =
\left\{
\begin{array}{ll}
     \eta_k (t)
     \,\hat{a} _k( t _0)+\hat{B}_k ^{(i)} (t)
     &\ \mathrm{if}\ i = 1,\\[1ex]
     \hat{B}_k ^{(i)} (t)
     &\ \mathrm{otherwise},
\end{array}
\right.
\end{equation}
where
\begin{equation}
\label{6.25} \hat{B}_k^{(i)}(t) = \int_{(\Delta_k)} \D\omega\,
\phi_{\mathrm {out}} ^{(i)}(\omega,t)\hat{B}_k(\omega,t)
,
\end{equation}
and it can be proved that
the commutation relations
\begin{equation}
  \label{7.7}
   \bigl[\hat{ a}_k  ( t _0),
   \hat{B}_k^{(1)\dagger} (t)\bigr] = 0
\qquad(t \ge t_0),
\end{equation}
\begin{equation}
  \label{7.13}
   \bigl[
    \hat{B}_k ^{(1)} (t),\hat{B}_k ^{(1) \dagger} (t)
   \bigr] = 1- \eta_k^2 (t)
\end{equation}
hold.

It is now straightforward to calculate
the characteristic function
$C^{(1)}_{k{\rm out}}(\beta, t)$ of the
quantum state of the relevant outgoing mode;
for details, see reference \cite{khanbekyan:043807}.
The result reads [$\sigma$ $\!=$ $\!(\mathrm{in},\lambda)$]
\begin{eqnarray}
    \label{7.16}
    C^{(1)}_{k{\rm out}}(\beta
    , t)
         &=&
         \exp
         \biggl
         [ - {\textstyle\frac{1}{2}} |\beta
           |^2
           [1-\eta ^2(t) - \sum _{\sigma , i} |\chi^{(i)}_{k\sigma}(t) |^2]
          \biggr
          ]
\nonumber\\
    &&\times\,
    C_{k}[\eta _k (t)\beta
    ]
    \prod _{
    \sigma
    i}
     C^{(i)}_{k
    \sigma
     }[\chi ^{(i)}_{k
    \sigma
     } (t)\beta
     ],
\end{eqnarray}
where the mode couplings are defined as
\begin{equation}
  \label{7.17}
    \chi^{(i)}_{k\sigma}(t)
    =
    \int_{(\Delta_k)} \D\omega\,
    \int_{(\Delta_k)} \D\omega'\,
    G_{k\sigma}
        (\omega, \omega ', t)
        \phi _{\mathrm {out}} ^{(1)*} (\omega, t)
        \phi _{\sigma} ^{(i)} (\omega', t_0),
\end{equation}
$C_{k}(\beta)$ is the characteristic function of the initially
prepared quantum state of the cavity mode,
and $C^{(i)}_{k\sigma }(\beta)$ are the
characteristic functions of the modes
of the incoming field
\mbox{($\sigma$ $\!=$ $\!\mathrm{in}$)} and the dissipative channels
($\sigma$ $\!=$ $\!\lambda$).
{F}rom equation (\ref{7.16}) the corresponding relation between
the Wigner functions is derived to be
[$\eta(t)$ $\!\equiv$ $\!\eta_k(t)$, $\chi_\sigma^{(i)}(t)$
$\!\equiv$ $\!\chi^{(i)}_{k\sigma}(t)$]
\begin{eqnarray}
\label{7.23}
\fl        W_{{\rm out}}( \alpha,t )
        = \frac {2} {\pi}
        \frac {1} {1 -\eta^2(t) - \sum _{\sigma i}
           |\chi_{\sigma } ^{(i)}(t)|^2
           }
\nonumber\\
\fl
\hspace{4ex}\times\,
     \int  \D ^2 \alpha ' \, W
        (\alpha ')
        \prod _{\sigma i}
        \int
        \D  ^{2} \alpha _{\sigma i}\,
        W_{\sigma }^{(i)}(\alpha _{\sigma i})
        \exp\!\left[ -\frac {2|
            \eta(t)
            \, \alpha '
            + \sum_{\sigma i }
            \chi_{\sigma  }  ^{(i)}(t)
        \,
        \alpha _{\sigma i}
        - \alpha|^2 }
        {1-\eta^2(t)-
        \sum _{\sigma i }
         |\chi_{\sigma }  ^{(i)}(t)|^2
  }     \right]
,
\end{eqnarray}
where ($\gamma_\mathrm{rad}$ $\!\equiv$
$\!\gamma_{k\mathrm{rad}}$, $\gamma_\mathrm{abs}$
$\!\equiv$ $\!\gamma_{k\mathrm{abs}}$)
\begin{equation}
  \label{8.4}
     \eta^2(t)  = \frac {\gamma_\mathrm{rad} ^{({\rm o})}}
        {\gamma_\mathrm{rad} + \gamma_\mathrm{abs}
        }
        \left[ 1- e^{-
        (\gamma_\mathrm{rad} + \gamma_\mathrm{abs})
        (t+ \Delta t-t_0)}\right],
\end{equation}
with
\begin{equation}
\label{8.6}
\gamma_{\mathrm{rad}}^{({\rm o})}
=
\gamma_{k\mathrm{rad}}^{({\rm o})}
     = \frac{c}{2 |n_1| l}   |T^{({\rm o})} _k|^2 .
\end{equation}

In particular, when the incoming modes and the
dissipative channels are in thermal states (with average number of
thermal photons $\bar{n}_ {\sigma i}$),
equation (\ref{7.23}) reduces to
\begin{eqnarray}
\label{8.3}
      W_{{\rm out}}(\alpha,t)
        = \frac {2} {\pi}\frac {
        1
        } {1-
        \eta^2(t)
        + 2
        \sum_ {\sigma i}
        \bar{n} _{\sigma i}
        |\chi _{\sigma}^{(i)}(t)|^2
        }
\nonumber\\\times\,
        \int \D ^2 \alpha'\,
        \exp\!\left[\! -\frac {2|
        \eta(t)
        \, \alpha'  -\alpha|^2 }
        {1-
        \eta^2(t)
        + 2
        \sum_ {\sigma i}\!
        \bar{n} _{\sigma i}
        |\chi _{\sigma}^{(i)} (t)|^2
        } \right]
       \!
        W
        (\alpha')
,
\end{eqnarray}
which reveals that the condition
\begin{equation}
\label{8.5}
        \frac{
        \eta^2(t)
        }{1-
        \eta^2(t)
        + 2
        \sum_ {\sigma i}
        \bar{n} _{\sigma i} |\chi _{\sigma } ^{(i)}(t)|^2
        } \gg 1
\end{equation}
must be satisfied to
ensure that the quantum state of the outgoing field becomes
sufficiently close to the quantum state that the cavity field was
initially prepared in.
It should be pointed out that the additional noise terms
in the input-output relation (\ref{2.87}), which are
associated with the losses in the coupling mirror, modify
the functions $G_{k\sigma}(\omega,\omega',t)$ [equations
(\ref{3.29})--(\ref{3.33})], which leads, compared to
standard noise theory, to an increased value of
$2\sum_{\sigma i}\bar{n}_{\sigma i}|\chi_\sigma^{(i)}(t)|^2$
in equation (\ref{8.3}), with the result that the condition of almost
perfect quantum state extraction becomes more restrictive.

\begin{figure}
\begin{center}
\includegraphics{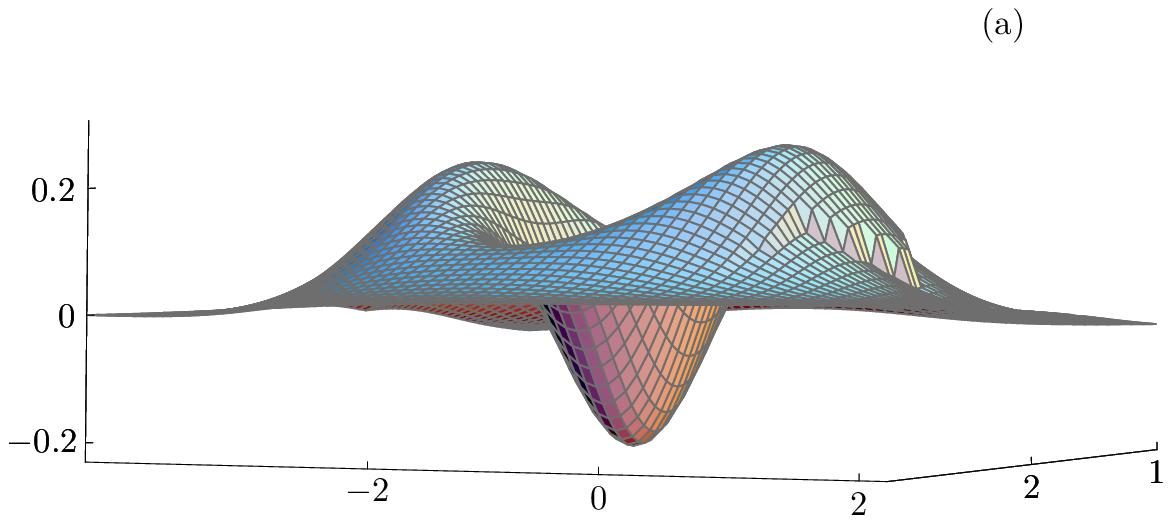}
 \vspace{8mm}\\
\includegraphics{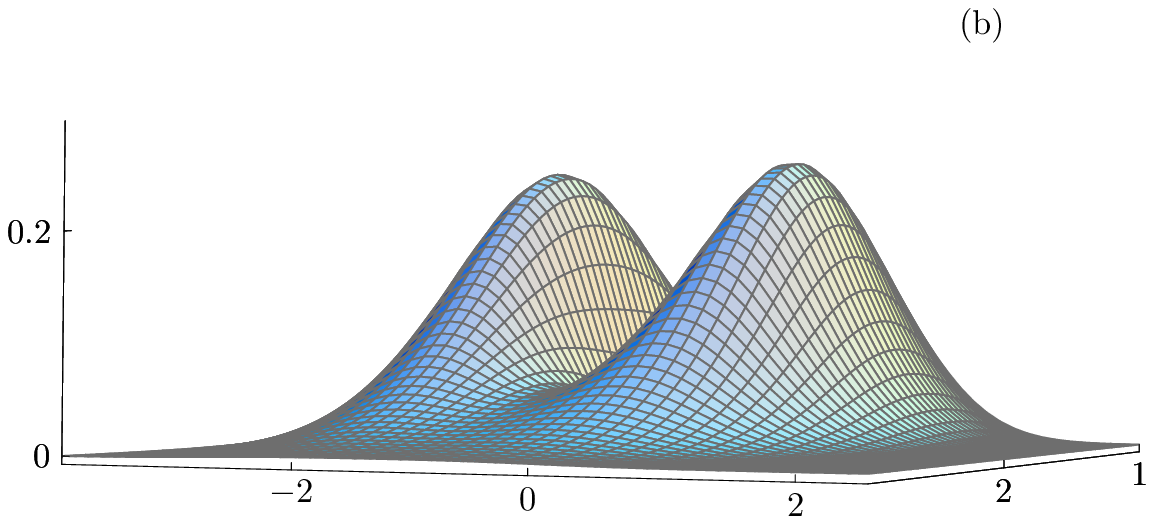}
\end{center}
\caption{\label{fid-1}
Wigner function of the
output state
for $n=1$,
$r=0.8$, $\bar{n} = 0.001$, $\beta = 2$,
(a)
$\chi = 0.1$,
$\chi _\mathrm{in}
= 0.9$,  $\eta
= 0.9$,
(b)
$\chi = 0.7$,  $\chi _\mathrm{in}
= 0.7$, $\eta
= 0.7$.
}
\end{figure}

\section{Generation of Schr\"{o}dinger cat-like states}
   \label{sec:6}

\begin{figure}
\begin{center}
\includegraphics{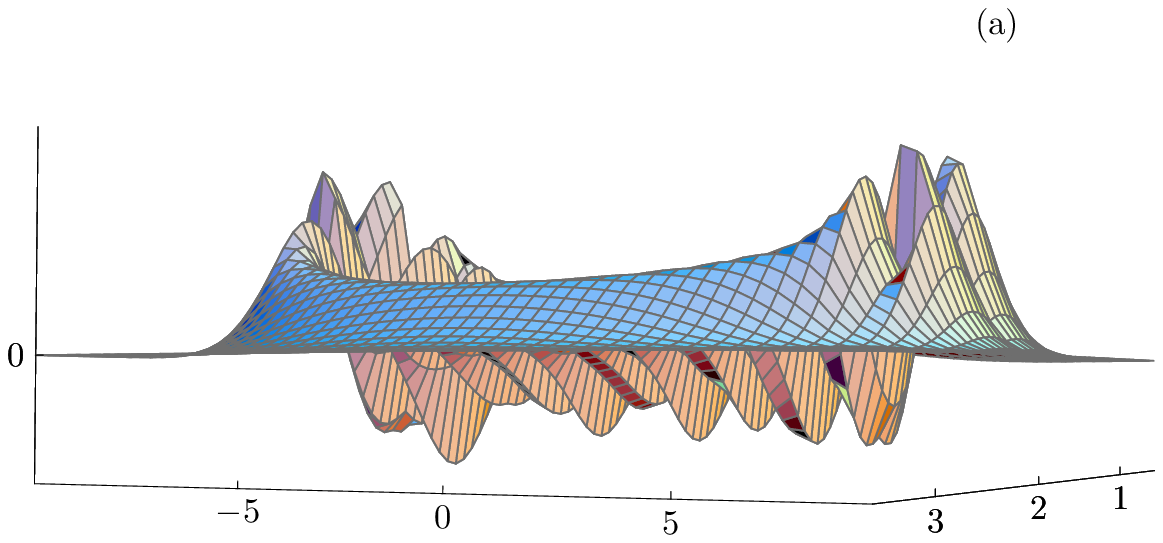}
 \vspace{8mm}\\
\includegraphics{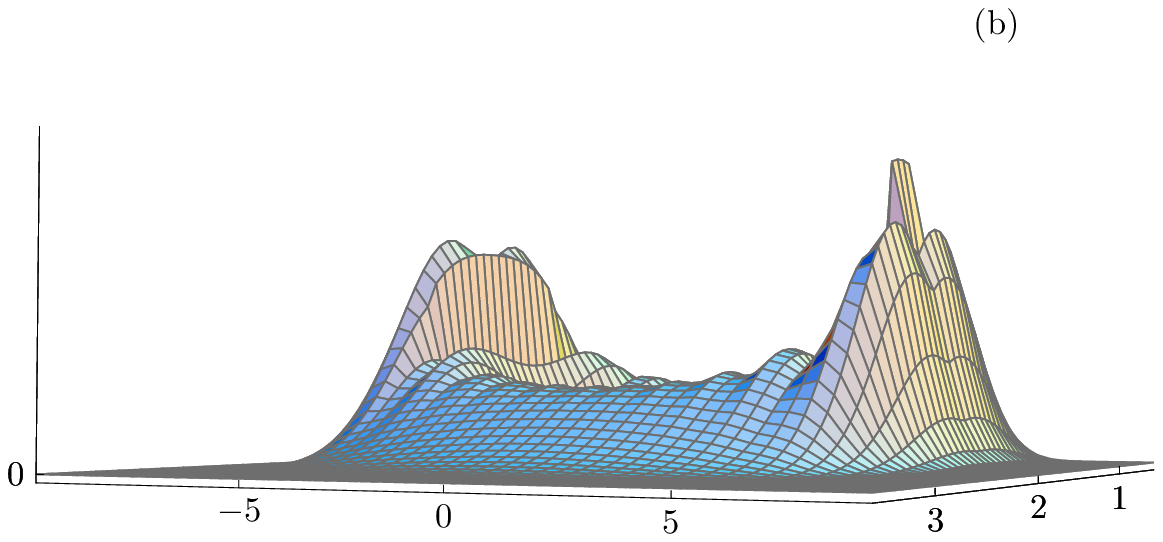}
\end{center}
\caption{\label{fid-2}
Wigner function
of the
output state
for
$n=10$,
$r=1.1$, $\bar{n} = 0.001$, $\beta = 2$,
(a)
$\chi = 0.1$,
$\chi _\mathrm{in}
= 0.9$,  $\eta
= 0.9$,
(b)
$\chi = 0.7$,  $\chi _\mathrm{in}
= 0.7$, $\eta
= 0.7$.
}
\end{figure}

As an application of the theory,
let us consider the case where two incoming modes
with $\chi^{(1)}_{\mathrm{in}}(t)$
$\!=$ $\!\chi^{(2)}_{\mathrm{in}}(t)
\equiv\chi_{\mathrm{in}}(t)$
are initially prepared in coherent states
$\left|\beta \right\ket $ and $\left|-i \beta \right\ket $,
respectively, and all other input modes and the
dissipative channels  are in thermal states with average numbers of
thermal quanta $\bar{n}$. Then, from equation~(\ref{7.23}) we find
\begin{eqnarray}
\label{9.0}
        W_{{\rm out}}(\alpha,t)
        = \frac {2} {\pi}\frac {
        1
        } {1-
        \eta^2(t)
        + 2
         \bar{n}
         |\chi (t)|^2
        }
\nonumber\\\hspace{2ex}\times\,
        \int \D ^2 \alpha'\,
        \exp\!\left[\! -\frac {2|
        \eta(t)
        \, \alpha'  -\alpha|^2 }
        {1-
        \eta^2(t)
        + 2
         \bar{n}
         |\chi (t)|^2
        } \right]
       \!
        W
        \Biggl
         [\alpha' -
        \frac{\beta\chi _\mathrm{in} ^{(1)} (t)}
        {
          \eta (t)
        }
        (1-i)
       \Biggr
       ]
,
\end{eqnarray}
where
the notation
\begin{equation}
  \label{9.3}
 {\sum_{\sigma i}} ^{\prime}
  \bar{n}
  |\chi_{\sigma }^{(j)} (t)|^2 =
  \bar{n}|\chi (t)|^2
\end{equation}
has been introduced, with the two incoming modes being excluded in
the primed sum. Let us suppose that the cavity mode is initially
prepared in a squeezed number state $| r, n\rangle$ $\!=$
$\!\hat{S} (r)| n\rangle$ [$\hat{S}(r)$, squeeze operator].
Examples of the Wigner function of the quantum state of the
outgoing mode are plotted in figures \ref{fid-1} and \ref{fid-2}
for $r = 0.8$, $n = 1$ and $r = 1.1$, $n = 10$, respectively, with
 $\bar{n} = 0.001$, $\beta = 2$ for both cases.
As we can see from figures \ref{fid-1}(a) and \ref{fid-2}(a), the
quantum state of the outgoing mode exhibits the typical features
of a Schr\"{o}dinger cat-like state, provided that the mode
coupling is strong enough ($\chi_\mathrm{in}\to 1$) and the noise
associated with the losses is sufficiently small ($\eta\to 1$,
$\chi\to 0$). Figures \ref{fid-1}(b) and \ref{fid-2}(b) reveal,
that with decreasing strength of the mode coupling and/or
increasing noise the nonclassical features of the quantum state of
the outgoing mode are completely lost. For the parameters chosen
the nonclassical oscillatory fringes typical for a Schr\"{o}dinger
cat-like state can be observed as long as $\eta(t)$ $\!\geq$
$\!0.7$, which for \mbox{$t$ $\!\to$ $\!\infty$}
 corresponds to the requirement that
$\gamma_\mathrm{abs}$ $ \lesssim$ $\gamma_\mathrm{rad}$. Note that
the additional noise terms in the input-output relation
(\ref{2.87}), which are associated with the losses in the coupling
mirror, effectively reduce the strength of the mode coupling
 $\chi_\mathrm{in}$. However, the necessary mode analysis
will be considered in detail in a forthcoming paper.

\section{Summary}
\label{sec:7}

To summarize, we have studied the
input-output problem for a high-$Q$ cavity,
with special emphasis on used input channels, taking into
account the absorption losses in the coupling mirror.
Within the framework of exact quantization of the
electromagnetic field in dispersing and absorbing media
we have performed the calculations for
a one-dimensional cavity bounded by a perfectly reflecting mirror and
a semi-transparent mirror.
The theory generalizes
the standard  quantum noise theory based on
a Hamiltonian of Senitzky-Gardiner-Collett type
\cite{senitzky:227,gardiner:1386},
which does not fully take into account
the absorption losses in the coupling mirror.
Using the generalized
operator input-output relations,
we have calculated
input-output relations for the Wigner functions of the
quantum states involved.
Finally, we have applied the theory to the case where two incoming modes
prepared in coherent states are combined with a cavity mode initially
prepared in a squeezed number state to prepare the outgoing field in
a Schr\"{o}dinger cat-like state.

\ack
This work was supported by the Deutsche Forschungsgemeinschaft.

\end{document}